\documentclass[letterpaper]{article} 
\usepackage[preprint]{aaai2027}  
\usepackage[hyphens]{url}  
\usepackage{graphicx} 
\usepackage{natbib}  
\usepackage{caption} 
\usepackage{algorithm}
\usepackage{algorithmic}

\usepackage{newfloat}
\usepackage{listings}
\DeclareCaptionStyle{ruled}{labelfont=normalfont,labelsep=colon,strut=off} 
\floatstyle{ruled}
\newfloat{listing}{tb}{lst}{}
\floatname{listing}{Listing}
\title{Learning Context-Free Grammars for Grammar-Constrained Decoding via Declarative Agentic Programming with Guarantees}

\author {
    Kevin Cheang\textsuperscript{\rm 1},
    Geoff Hulette\textsuperscript{\rm 1},
    Rahul Kumar\textsuperscript{\rm 1},
    Felipe R. Monteiro\textsuperscript{\rm 1}, \\
    Federico Mora\textsuperscript{\rm 1},
    Robin Salkeld\textsuperscript{\rm 1},
    Lin Tan\textsuperscript{\rm 1, \rm 2},
    Serdar Tasiran\textsuperscript{\rm 1}
}
\affiliations {
    \textsuperscript{\rm 1}Amazon Web Services\\
    \textsuperscript{\rm 2}Purdue University
}

\usepackage[utf8]{inputenc}
\usepackage{booktabs}
\usepackage{multirow}
\usepackage{amsfonts}
\usepackage{amsmath}
\usepackage{nicefrac}
\usepackage{microtype}
\usepackage{xcolor}
\usepackage{subcaption}
\usepackage{soul}
\usepackage{tikz}
\usetikzlibrary{positioning, arrows.meta, calc, fit}
\usepackage{pifont}
  
\newcommand{\AG}{Autogrammar}

\newtheorem{example}{Example}

\begin{document}

\maketitle

\begin{abstract}
Language models (LMs) are increasingly used to interact with external services via programs written in domain-specific languages (DSLs). Unfortunately, since DSLs are often low-resource and esoteric, LMs frequently produce syntactically invalid programs in these languages. Grammar-constrained decoding can eliminate such failures, but requires syntactic constraints. These are usually in the form of a context-free grammar for the target language, an artifact that is hard to come by for third-party DSLs. In this work, we define an agent, called \AG{}, that automatically learns context-free grammars from documentation and execution data. \AG{} is formalized as a Kripke structure whose nondeterministic choices are resolved by a language model, enabling declarative control of agent behavior via linear temporal logic constraints. We evaluate four versions of \AG{} on three DSLs (i.e., Amazon CloudWatch Logs Insights, Dynatrace Query Language, and Datadog Search Syntax) and find that it generates grammars that achieve near perfect precision on unseen data; that temporal restrictions reduce execution time by $3.8\times$ without incurring statistically-significant loss in precision; that execution data is crucial while documentation is dispensable; and that grammar-constrained decoding using \AG{}-generated grammars significantly improves end-to-end LM performance on eight out of ten real tasks, matching or exceeding the performance of a professionally-maintained grammar. In comparison, the context-free grammars generated by existing LM baselines and a state-of-the-art formal technique perform significantly worse over the same evaluation.
\end{abstract}


\section{Introduction and Motivation}

Language models (LMs) are increasingly used to generate code in domain-specific languages (DSLs) that have far less training data than general-purpose programming languages.
For example, \citet{11105878} use LMs to generate Verilog designs; \citet{NEURIPS202372223cc6} to generate SQL queries; and \citet{NEURIPS202444af0654} to generate PDDL plans. The cloud domain in particular is replete with DSLs. Using Claude Sonnet 4.6 to classify every string-typed parameter across 416 Amazon Web Services (AWS) API services available in the Boto3 Python SDK, we found that 23\% of services expose a string parameter that must conform to a DSL. In total, we found that AWS uses at least 59 unique DSLs,\footnote{One author manually confirmed that all 59 unique DSLs correctly correspond to at least one string-typed parameter.} including policy languages, like IAM policies, query languages, like SQL dialects, and batch scheduling languages, like cron. For example, the \texttt{Query} string parameter for the \texttt{StartQuery} API operation\footnote{See \url{https://docs.aws.amazon.com/AmazonCloudWatchLogs/latest/APIReference/API_StartQuery.html}.} requires a DSL program, while the \texttt{LogGroupName} string parameter does not. 

\begin{figure*}[t]
  \centering
  \resizebox{\textwidth}{!}{\input{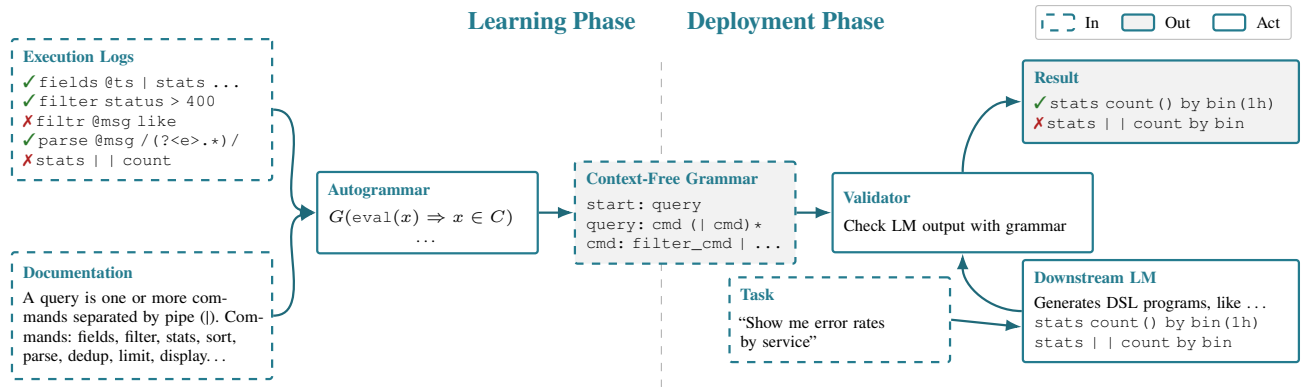}}
  \caption{\AG{}'s two-phase pipeline. In the \textbf{learning phase}, \AG{} consumes execution logs (labeled valid/invalid programs) and API documentation to produce a context-free grammar for the target DSL. In the \textbf{deployment phase}, a downstream LM uses the context-free grammar through a validator to generate correct DSL programs in response to natural-language tasks.}
  \label{fig:overview}
\end{figure*}

Unfortunately, many studies show that LM performance degrades sharply on low-resource DSLs. For example, in a recent survey, \citet{sathvik2026} find that LM output quality across programming languages correlates with training data availability. None of the 59 unique DSLs that we found across AWS are as high-resource as general-purpose languages, like Python. Therefore, it is reasonable to expect suboptimal out-of-the-box LM performance across API operations in 23\% of AWS services. We empirically confirm this for one AWS DSL in Section~\ref{sec:evaluation}.

Grammar-constrained decoding (GCD) addresses this problem by restricting LM outputs to strings that conform to a syntactic specification. For example, \citet{scholak2021picard} introduce incremental parsing alongside the decoder to reject tokens that cannot extend to a syntactically valid prefix; \citet{geng2023gcd} show that off-the-shelf LMs with input-dependent grammars match task-specific fine-tuned models on structured NLP tasks. These approaches can be made to scale and preserve the semantic prowess of LMs. For example, \citet{willard2023efficient} cast GCD in terms of finite-state machine transitions to improve decoding performance; \citet{park2024gad} give an algorithm that produces outputs proportional to the LM distribution conditioned on an input grammar. We provide a comprehensive survey of GCD and related techniques in Section~\ref{sec:related}.

Regardless of the theoretical variations or implementation details, all of these approaches require syntactic constraints as input, frequently in the form of a context-free grammar. But where does one obtain a grammar for GCD? For well-established languages, hand-written grammars may already exist. For the long tail of DSLs---proprietary query languages, vendor-specific syntax, evolving configuration formats---no grammar is publicly available, and writing one by hand is expensive and error-prone. For example, \citet{10.1109/ASE.2019.00047} empirically demonstrate the challenges of manually writing regular expressions, which correspond to a restricted form of context-free grammars. This motivates the central challenge of this paper: \emph{before} we can use constrained decoding tools, we must first learn a high-quality grammar.

Grammar inference is a classical problem in formal language theory, dating to at least as far back as Gold's seminal work on context-free grammar identification in the limit~\cite{gold1967language}. Subsequent decades produced algorithms for learning regular languages, such as $L^*$~\cite{angluin1987learning}, and restricted subclasses of context-free languages that can be learned from examples~\cite{clark2007polynomial}. A systems-oriented line of work eschews formal guarantees for practical techniques that learn useful context-free grammars \cite{bastani2017glade,kulkarni2022arvada}. However, these methods assume access to a membership oracle that can be queried on arbitrary strings. This is a strict assumption when working with DSLs that can only be accessed behind authenticated, expensive, rate-limited APIs that might mutate resources.

In this paper, we propose and evaluate an approach that takes inspiration from this history but uses modern tools to solve the problem in a new setting. Specifically, we design a grammar-learning agent, called \AG{}, that passively learns context-free grammars from documentation and execution data. \AG{} is neither a fixed algorithmic procedure, like classic grammar learning techniques, nor a general-purpose reasoner, like modern coding agents. Instead, we formalize \AG{} as a Kripke structure whose non-deterministic choices are resolved by an LM and whose behavior is constrained by linear temporal logic (LTL) constraints. This framing allows us to quickly declare, in the sense of the declarative programming paradigm, different versions of our agent for different settings. For example, in Section~\ref{sec:evaluation}, we show that by only changing LTL constraints, we can reduce execution time by $3.8\times$ while still generating context-free grammars that achieve near perfect precision on unseen data. Our approach, therefore, balances the power of LMs with the precise control of fixed algorithmic procedures.


Figure~\ref{fig:overview} illustrates our approach. During the learning phase, \AG{} ingests execution logs together with available documentation for the target DSL. We assume that these execution logs come from an existing LM agent that is actively struggling to write DSL code. If the existing agent were not struggling to write DSL code, developers would not need to turn to GCD. \AG{} then iteratively proposes, evaluates, and refines candidate grammars until it either finds a context-free grammar that accepts the valid examples and rejects the invalid ones, or hits a resource limit. This iterative process is constrained and defined by LTL constraints. During the deployment phase, the learned grammar is handed to an implementation of GCD. We call this the ``Validator,'' but it can be any existing GCD tool that uses context-free grammars. At this point, the original LM agent that struggled to write DSL code, labeled ``Downstream LM'' in Figure~\ref{fig:overview}, works with the validator to produce correct DSL programs. Overall, our approach and this paper make the following main contributions: 
\begin{itemize}

\item \textbf{Declarative Agent Programming Framework}. We frame LM agents as Kripke structures whose non-deterministic choices are resolved by LMs. This framing enables the declarative definition of agent behavior via LTL constraints. We demonstrate the framework by defining four versions of a new agent, \AG{}, that learns context-free grammars for DSLs from execution data and documentation. All versions are identical but for the LTL constraints that control the agent's behavior. In Section~\ref{sec:evaluation}, we show that these simple high-level edits can reduce \AG{}'s execution time by $3.8\times$ while still having it generate context-free grammars that achieve near perfect precision on unseen data.

\item \textbf{Significantly Better Context-Free Grammars}. We use \AG{} to generate context-free grammars for three DSLs (CloudWatch Logs Insights, Dynatrace Query Language, and Datadog Search Syntax) and show that \AG{}-generated grammars achieve an average of 91--100\% precision on unseen data over a five-fold cross-validation. These results are significantly better than those obtained by existing LM baselines and a state-of-the-art formal technique for learning context-free grammars from data. We also find that execution data is crucial while documentation is largely dispensable.

\item \textbf{End-to-End Improvements with GCD Integration}. We integrate \AG{}-generated grammars into a simple GCD pipeline and show that this significantly improves end-to-end LM performance on 80\% of tasks in a new benchmark. The benchmark consists of 10 Amazon CloudWatch Logs Insights tasks that were used to analyze executions of \AG{} itself. These tasks are real and contributed to the empirical analysis in this paper. Furthermore, we show that, on these tasks, \AG{}-generated grammars match or exceed the performance of a professionally maintained grammar for the same DSL.

\end{itemize}

\section{Background Needed for Approach}
In this section, we give the minimal background on Kripke structures and linear temporal logic (LTL) required to understand our approach. For a comprehensive background on both concepts, see \citet{clarke1997model}. For a background on context-free grammars---the artifacts that we generate---see \citet{lucas2026seventy}.

A Kripke structure is a 5-tuple, $(S, I, R, L, {AP})$, where $S$ is a (possibly infinite) set of states, $I \subseteq S$ is a set of initial states, $R \subseteq S \times S$ is a transition relation, $L : S \to 2^{AP}$ is a labeling function, and ${AP}$ is a set of atomic propositions over states. We often define the set of states with typed variable declarations; the set of initial states as a predicate over those typed variables; and the transition relation as a predicate over two copies of those typed variables. 

\begin{example}[Coin Flipping Kripke Structure]
\label{ex:coin}
Consider a system that generates a random bit string by repeatedly flipping a coin. We can model this as a Kripke structure whose set of states, $S$, is defined by two typed variables \texttt{coin: bool} and \texttt{y: string}; whose set of initial states, $I$, is defined by the predicate \texttt{y == ""}; and whose transition relation is defined by the predicate \texttt{y' = y + "1" if coin else y + "0"}. 
\end{example}

In the transition relation above, \texttt{y'} refers to the next value of the variable and \texttt{y}, without the tick, refers to the current value. The transition relation predicate constrains the values that variables can take on over time. Since \texttt{coin'} does not appear in the predicate for our example above, the value that \texttt{coin} takes on next is unconstrained. 

The labeling function $L$ and the set of propositions ${AP}$ can be used to constrain the behavior of the system over time using LTL. For example, we can use LTL to enforce that it is always the case that eventually our random bit string does not end in a repeated bit. Let $p \in {AP}$ be the proposition that the state variable $y$ does end in a repeated bit: \texttt{p $\doteq$ y[-1] == y[-2]}. The formula $GF\neg p$ formally captures our constraint. In LTL, $Gp$ means that $p$ always (``G'' is for ``globally'') holds; $Fp$ means that $p$ eventually (``F'' is for ``finally'') holds. These temporal operators, along with ``X'' (for ``holds in the \emph{next} step''), ``U'' (for ``until''), and ``W'' (for ``until without the requirement that it eventually holds'') can be nested and combined with boolean operators to form constraints that enforce complex system behaviors. LTL is usually used to specify what a system \emph{should} do, not constrain what it does. In this paper, we use LTL to constrain what systems do.

\section{Approach: LTL Constrained Agents}
\label{sec:approach}

In this section, we describe a new approach for defining LM agents that are parameterized by LTL constraints, we define one parametric LM agent called \AG{}, and we create four instances of \AG{} using four different sets of LTL constraints. \AG{} learns context-free grammars from documentation and execution data. All four versions of \AG{} include safety constraints, like that the agent never uses the testing data during training, but differ in the constraints which, informally speaking, encode learning strategies. In Section~\ref{sec:evaluation} we show that encoding learning strategies as LTL constraints can lead to significant performance improvements while preserving rigorous guarantees. Overall, we argue that our approach provides developers with a new way to engineer safe and performant LM agents.

The key idea behind our approach is to define parametric LM agents as Kripke structures where, during execution, all non-deterministic choices are made by querying an LM. The definition of the Kripke structure itself limits the choices that the LM can make. For example, the coin flipping Kripke structure in Example~\ref{ex:coin} only allows one non-deterministic choice: at each step, the coin can be heads (\texttt{True}) or tails (\texttt{False}). We use LTL constraints to further restrict the non-deterministic choices. For example, \texttt{G$\neg$(y[-1] == y[-2] == y[-3])} enforces that there will never be three consecutive heads or tails, assuming \texttt{y} is longer than three characters. If the current state of the bit string \texttt{y} is \texttt{"011"}, then the aforementioned LTL constraint will disallow a value of \texttt{True} for \texttt{coin}. In practice, if the underlying LM attempts to make a non-deterministic choice that would violate an LTL constraint, then our implementation would catch this and ask the LM to try again. In other words, we implement these constraints through rejection sampling.

\subsection{\AG{} as a Kripke Structure}

We now formally define \AG{} as a Kripke structure.

\paragraph{S, I, and R} \AG{} consists of state variables that hold training data (positive and negative string examples), the documentation for the target DSL, candidate context-free grammars, and metadata, like the minimum number of correct rejections required and a timestamp marking when the agent started execution.
Initially, the training data comes from previously collected execution logs and the set of candidate context-free grammars is empty. 
At every step, the LM powering \AG{} picks between a fixed set of actions that allow it to interact with the environment. These include:
\begin{enumerate}
    \item \texttt{add\_g} --- Propose a candidate context-free grammar.
    \item \texttt{eval} --- Test a grammar against all training examples.
    \item \texttt{add\_s} / \texttt{del\_s} --- Add/remove from the training set.
    \item \texttt{gen\_ts} --- Use documentation to generate strings targeting flaws in a grammar and add to the training set.
    \item \texttt{return} --- Return the final grammar.
\end{enumerate}
In short, the transition relation is a nondeterministic choice between a set of deterministic tools that operate on the state.

\paragraph{L and AP} Every time a tool is called, the call, including its arguments, is written to a state variable called \texttt{head}. For example, if there are two tools, $f$ and $g$, both of which take a single integer argument, then $f(0)$, $g(1)$, and $f(100)$ are all values that \texttt{head} could take on during an execution. The \texttt{head} variable is a \emph{ghost variable}: no action modifies it directly and it is only used for defining labels. \AG{}'s labels represent predicates over the \texttt{head} variable and the other state variables. For example, in \AG{}'s setting, evaluating a grammar is time consuming, so we will want to specify that no grammar is evaluated twice without changing the underlying data in between. To do this, we use predicates that capture the value of \texttt{head}. For example, we use \texttt{eval(x)} to mean that the given state has \texttt{head = eval(x)}, for some $x$ from the set of candidate context-free grammars. Given these predicates and corresponding labeling function, we enforce the desired behavior using
\begin{align}
\label{eq:efficient-ltl}
\forall x.\; G\bigl(\texttt{eval}(x) \implies X\bigl(G \neg\texttt{eval}(x) \notag\\
  \;W\, \exists y\, (\texttt{add\_s}(y) \lor \texttt{del\_s}(y))\bigr)\bigr),
\end{align}
which reads ``for all $x$, globally, if \texttt{eval}($x$) holds, then from the next state onward, it is always the case that \texttt{eval}($x$) does not hold until there is some $y$ for which \texttt{add\_s}($y$) or \texttt{del\_s}($y$) hold.''

\paragraph{LTL} In Section~\ref{sec:evaluation} we compare four instances of \AG{}. All four versions are exactly the same except for the LTL constraints that we impose on them. In this way, \AG{} is parametric. The version called ``Unlimited'' has the fewest LTL restrictions: the underlying LM is mostly allowed to make any nondeterministic choices. The few restrictions are LTL constraints like 
\begin{equation}
\forall x.\; G (\texttt{eval}(x) \implies x \in C), 
\end{equation}
which reads ``for all $x$, globally, whenever \texttt{eval}(x) is called, $x$ must be a member of $C$,'' where $C$ is the set of candidate context-free grammars. The version called ``Performance'' adds efficiency constraints, like Equation~\ref{eq:efficient-ltl}. It also adds a strict bound on the amount of time that the agent can take. Let $t_0$ be the time at which the execution started, $b$ be the maximum amount of time given to the agent, and $t$ be the current time. For ``Performance'' we enforce that
\begin{equation}
\label{eq:bound}
G ((t - t_0 \geq b) \implies X \; \exists x.\; \texttt{return}(x)).
\end{equation}
The versions called ``No Documentation'' and ``No Execution Data'' extend ``Performance'' by disallowing actions that read documentation or execution data, respectively. 

\paragraph{Formal Guarantees}
The structure and semantics of \AG{} can be used to provide formal guarantees about its execution. The two most important guarantees relate to \emph{safety} and \emph{progress}. For safety, \AG{} ensures that the underlying LM cannot train on testing data. This follows from the set of available actions---the tools that interact with execution data, \texttt{read} and \texttt{eval}, only access training data. General-purpose programming agents, like Claude Code and Copilot, do not provide similar guarantees without more heavyweight user interventions, like sand-boxing. Our Kripke structure makes these guarantees explicit and verifiable. On the other hand, Equation~\ref{eq:efficient-ltl} and similar constraints encode a notion of progress. For example, test generation is expensive, so we enforce that it is never repeated on the same grammar:
\begin{equation}
\label{eq:no-repeat-gen}
\forall x.\; G\bigl(\texttt{gen\_ts}(x) \implies X\; G \neg\texttt{gen\_ts}(x)\bigr),
\end{equation}
which reads ``for all $x$, globally, if \texttt{gen\_ts}($x$) holds, then from the next state onward, \texttt{gen\_ts}($x$) never holds again.'' Recent studies show that LMs can get caught in unproductive loops \cite{li2026earlydiagnosiswastedcomputation,Adnan2025}. We use LTL to automatically prevent specific unproductive patterns that we observed in early prototypes.

\subsection{Generic Agent Programming}

Our approach mirrors how developers already build agents in practice. Frameworks like Strands allow developers to define an agent by specifying a set of tools---exactly the pattern we formalize as a Kripke structure with nondeterministic tool selection. What we add is a formal-methods framing: the Kripke structure makes the state space explicit, and LTL constraints provide declarative, verifiable restrictions on behavior. Because this formalization directly captures existing development practice, we believe it constitutes a natural programming model for constrained agents. This also suggests that the approach will generalize beyond learning context-free grammars to other agentic domains, though further empirical study is required to confirm this.
\section{Empirical Evaluation}
\label{sec:evaluation}

In this section, we empirically evaluate four versions of \AG{} and the
context-free grammars they generate. To do so, we implement a prototype in Python using approximately 2300 lines of code. We use the Strands SDK to interact with LMs and we use Lark to represent context-free grammars; we prompt an LM, through Strands, to generate a Lark grammar that achieves perfect precision on a training set of examples. We prioritize precision because it is compatible with the goals of GCD: to block incorrect syntax without blocking correct syntax. We conduct all experiments on an Amazon EC2 instance with 16 virtual CPUs, 32 GiB of memory, and up to 12500 Megabit network performance (c7a.4xlarge), seeking to answer the following research questions.
\begin{enumerate}
    \item Does \AG{} generate context-free grammars that achieve perfect
    precision on unseen data?
    \item How do \AG{}-generated context-free grammars compare to those generated by state-of-the-art formal techniques and LM baselines?
    \item Can \AG{} with performance constraints generate context-free grammars of
    similar quality using a fraction of the time?
    \item Does \AG{} need access to both documentation and execution data? If not, which input is more important?
    \item Can \AG{} grammars improve end-to-end performance compared to LMs without grammars or LMs using existing grammars?
\end{enumerate}

In the end, we find that \AG{} grammars achieve almost perfect precision on
unseen test data; that \AG{} with LTL performance constraints generates grammars that are statistically no different when it comes to grammar precision while using a fraction of the time; that \AG{} works well without documentation but that execution data is crucial
for its success; and that \AG{} grammars significantly improve end-to-end
LM performance, comparable to professionally maintained existing
grammars.

\subsection{RQ1-4: Grammar Analysis}

For the first four research questions, we use Claude Opus 4.6 and we focus on three
DSLs: Amazon CloudWatch Logs Insights Language (CWIL), Dynatrace
Query Language (DQL), and Datadog Search Syntax (DSS). For CWIL, we have access
to 41183 internal API calls, 1016 of which are negative examples and 40167 of
which are positive examples. For DQL, we have access to 1723 internal API calls,
354 of which are negative examples and 1369 of which are positive examples. For
DSS, we have access to 838 internal API calls, 17 of which are negative examples
and 821 of which are positive examples. For all three DSLs,
we use the official language documentation webpages as input to \AG{}.

\subsubsection{RQ1: Perfect Precision}

\begin{table}[t]
\centering
\caption{Precision and recall per DSL per variant (mean$\pm$std).}
\label{tab:variant-comparison}
\small
\resizebox{\columnwidth}{!}{%
\begin{tabular}{llccc}
\toprule
\textbf{Variant} & \textbf{Metric} & \textbf{CWIL} & \textbf{DQL} & \textbf{DSS} \\
\midrule
\multirow{2}{*}{Unlimited}
  & Prec. & .999 $\pm$ .003 & 1.00 $\pm$ .000 & .910 $\pm$ .124 \\
  & Rec.  & .555 $\pm$ .197 & .514 $\pm$ .060 & .950 $\pm$ .112 \\
\midrule
\multirow{2}{*}{Performance}
  & Prec. & .989 $\pm$ .025 & 1.00 $\pm$ .000 & .883 $\pm$ .162 \\
  & Rec.  & .176 $\pm$ .112 & .545 $\pm$ .064 & 1.00 $\pm$ .000 \\
\midrule
\multirow{2}{*}{No Documentation}
  & Prec. & .988 $\pm$ .018 & .975 $\pm$ .016 & .768 $\pm$ .215 \\
  & Rec.  & .584 $\pm$ .188 & .644 $\pm$ .041 & 1.00 $\pm$ .000 \\
\midrule
\multirow{2}{*}{No Execution Data}
  & Prec. & .113 $\pm$ .034 & .892 $\pm$ .196 & .770 $\pm$ .144 \\
  & Rec.  & .306 $\pm$ .104 & .477 $\pm$ .040 & .950 $\pm$ .112 \\
\bottomrule
\end{tabular}%
}
\end{table}

To answer the first research question, we conduct a 5-fold cross validation. For
each DSL, we split the data into five sets through
stratified sampling, ensuring proportional numbers of positive and negative
examples in each set. We then generate a context-free grammar using four of the
five sets, test the context-free grammar as a successful API call classifier on
the fifth set, report the test results, and then repeat the experiment with the
next combination. The ``Unlimited'' columns of Table~\ref{tab:variant-comparison} show the results. We see that \AG{} does
not generate grammars with perfect precision on the test set every time. However, for two of the
three case studies---the two with more available data---\textit{\AG{} generates
grammars that achieve almost perfect precision on unseen test data.}

\subsubsection{RQ2: Baseline Comparison}

To contextualize the results of RQ1, we compare \AG{} with Arvada~\citep{kulkarni2022arvada} 
and a baseline LM approach. Arvada is a state-of-the-art grammar induction algorithm 
that learns context-free grammars using an oracle. During its execution, Arvada generates strings that may belong to the target language; Arvada uses the oracle to determine if these new strings do belong to the target language. In this way, Arvada is an \emph{active learning} approach whereas \AG{} is passive. Active learning approaches require oracles but often outperform passive learning approaches~\cite{hanneke2025agnostic}.

We use Arvada to learn context-free grammars for CWIL over a 5-fold cross-validation and average the results. Whenever Arvada seeks to classify a candidate string, we use an existing grammar that is maintained by the CloudWatch team for query validation in the publicly available web frontend. The second row of Table~\ref{tab:baseline} contains the results of this experiment. With an unlimited oracle budget, Arvada makes an average of 63k oracle calls, takes an average of 14.6 hours per run, and generates context-free grammars that achieve an average precision of 0.025 on the unseen data per fold---the learned grammar rejects
97.5\% of valid queries. This is in stark contrast to \AG{}'s near perfect precision. We do not evaluate Arvada on DQL or DSS since we do not have access to a ground-truth grammar.

In other circumstances, Arvada performs exceptionally well \cite{kulkarni2022arvada}. The key issue in our setting is that the CloudWatch service is deliberately permissive. For example, it accepts arbitrary text as filter expressions (e.g., \texttt{"xyzzy foo bar"}). Arvada's
bubble-and-merge algorithm relies on negative examples from the oracle to discover
structure: when a string replacement is rejected, Arvada learns that two
substrings play different syntactic roles. Without sufficient negative examples, we find that
Arvada cannot generalize well beyond the initial training examples,
producing a grammar that is effectively a memorization of the input examples for CWIL. On the other hand, \AG{} leverages the underlying LM's pre-trained knowledge of programming languages to construct a context-free grammar, rather than relying on negative oracle responses to generalize.

But how important is the LM's pre-trained knowledge of programming languages? Can an LM generate a good context-free grammar without any of \AG{}'s contributions? To answer these questions, we implement two versions of an LM baseline: \emph{LM-0}, which prompts an LM to generate a context-free grammar from the same documentation that \AG{} reads; and \emph{LM-40}, which extends the LM-0 prompt with 20 positive and 20 negative examples. Both prompts open with the literal first sentence of \AG{}'s
agent prompt. Both versions of the LM baseline use Claude Opus 4.6, retry up to three times if the produced text fails to parse as a Lark grammar, and are evaluated in the same way as \AG{} was for RQ1. 

The third and fourth rows of Table~\ref{tab:baseline} show the results of our LM-baseline experiment.
LM-40 is the strongest of the two and reaches at most $0.51$
precision on DQL, $0.07$ on CWIL, and $0.12$ on DSS; an order of
magnitude below \AG{} on two of the three DSLs. Across all configurations that we evaluated, \AG{} dominates on precision under a two-proportion z-test. All comparisons survive Holm-Bonferroni correction at $\alpha=0.05$ to control the family-wise false-positive rate. Therefore, we conclude that the performance gap between \AG{} and Arvada is not
explained by the LM's pre-trained knowledge alone. \AG{}'s contributions are what turn documentation and execution data into a
near-perfect grammar. \textit{\AG{}-generated grammars significantly outperform those generated by state-of-the-art formal techniques and LM baselines.}

\begin{table}[ht]
\centering
\small
\resizebox{\columnwidth}{!}{%
\begin{tabular}{lcccc}
\toprule
Method & Oracle Calls & CWIL Prec. & DQL Prec. & DSS Prec. \\
\midrule
\AG{} & N/A & $.999 \pm .003$ & $1.00 \pm .000$ & $.910 \pm .124$ \\
Arvada & $63\text{K} \pm 12\text{K}$ & $.025 \pm .001$ & N/A & N/A \\
LM-0 & N/A & $.046 \pm .019$ & $.228 \pm .067$ & $.023 \pm .006$ \\
LM-40 & N/A & $.074 \pm .028$ & $.510 \pm .106$ & $.115 \pm .158$ \\
\bottomrule
\end{tabular}%
}
\caption{Generated grammar precision comparison. Only Arvada uses an oracle. Oracle only available for CWIL.}
\label{tab:baseline}
\end{table}

\subsubsection{RQ3: Performance}

In the previous experiment, \AG{} takes an average of 461 minutes to learn a
context-free grammar for CWIL. To answer RQ3, we use the performance constraints described in Section~\ref{sec:approach}. For example, we set $b$ in Equation~\ref{eq:bound} to 120 minutes. We then measure the degradation in precision and recall.
Table~\ref{tab:variant-comparison} shows the results along with additional data for RQ4. Specifically, the ``Performance'' columns show the precision and recall of \AG{} grammars per DSL under the 120 minute time limit.

Table~\ref{tab:variant-comparison} shows that precision does not decrease when performance constraints are added. In fact, \textit{there is no statistically significant
difference between \AG{} with and without them for any
DSL when it comes to precision.} On the other hand, recall is impacted. For CWIL, there is a statistically significant 
drop in recall with these constraints.

\subsubsection{RQ4: Ablation Study}

Table~\ref{tab:variant-comparison} contains two additional configurations 
that were not used to answer RQ3. The first, ``No Documentation,'' refers 
to a configuration of \AG{} without any action that depends on documentation. The 
second, ``No Execution Data,'' refers to a configuration of \AG{} without 
any action that depends on execution data. Both configurations include the 120 
minute bounds. \textit{Surprisingly, \AG{} works well without 
documentation for CWIL and DSS.} There is no statistically significant reduction in precision or 
recall when documentation is removed from these two DSLs---even with the bounds. For DQL, removing documentation leads to a small ($1.00$ to $0.975$) but statistically significant reduction in precision ($p=0.027$). 
This suggests that for some DSLs, the LM may not effectively leverage documentation. For CWIL and DSS, the
time it is spending reading documentation does not translate to an improvement
in the quality of the generated grammar.

On the other hand, \textit{execution data is crucial.} Without execution data,
\AG{} performs significantly worse in terms of precision and recall for CWIL.
Similar but less severe reductions in performance exist for DQL. DSS
performance is the least impacted, likely due to the noise introduced through
the lack of available data.

\subsection{RQ5: End-to-End Impact}

\begin{figure}[t]
    \centering
    \includegraphics[width=\columnwidth]{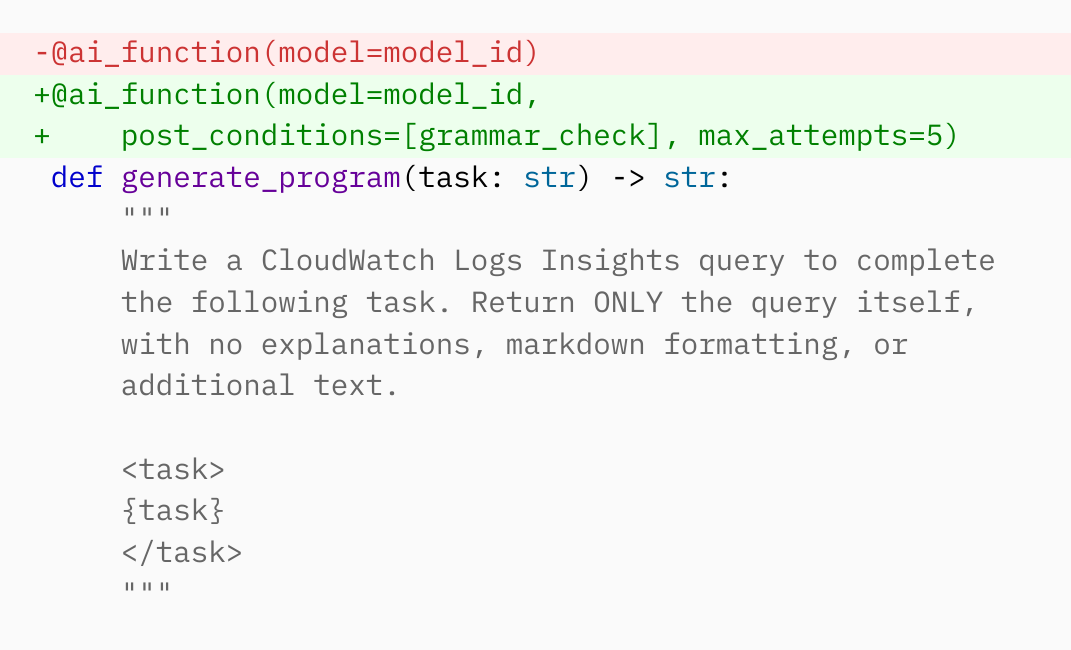}
    \caption{AI function for generating queries. The highlighted lines show the
    added post-condition.}
    \label{lst:aifunc}
\end{figure}

To answer the last research question, we instrument \AG{} so that it generates
CloudWatch Logs during its execution, we implement three different versions of a
CloudWatch Logs Insights query agent, and we use all three to solve 10 natural-language tasks. Each task corresponds to a real analysis of \AG{} used in the
writing of this paper or the development of \AG{} itself. For example, ``Task
10: Timestamp Arithmetic'' was used to collect data on execution times for
all the different versions of \AG{}.

All three agents are implemented as a Strands AI function displayed in
Figure~\ref{lst:aifunc}. The first version of our AI function does not use a
post-condition: we simply ask an LM to generate a query that solves a
given task. The second and third versions add a post-condition, shown as the
highlighted diff in Figure~\ref{lst:aifunc}. The post-condition is a predicate
called \texttt{grammar\_check} that takes a candidate query and checks it using
a context-free grammar. For the second version of our agent, we use a grammar
generated by running \AG{} on all available data for CWIL. For the third version
of our agent, we use an existing grammar that is maintained by the CloudWatch
team for query validation in the publicly available web frontend. This is the same grammar from RQ2.

When executing AI functions, the Strands runtime queries the underlying language
model repeatedly until the LM generates an output that passes the
post-condition check, or a hard iteration limit is reached (5 attempts in our
experiments). In our experiments, we use six different LMs
(claude-opus-4-6-v1, claude-sonnet-4-6, claude-haiku-4-5-20251001-v1:0,
nova-pro-v1:0, qwen3-32b-v1:0, deepseek.v3.2), and we call each LM
on each task 100 times. For each call, we run the generated CloudWatch Logs
Insights query through boto3 and check if the call succeeded (no errors
or exceptions).

\begin{figure*}[t]
    \centering
    \includegraphics[width=\textwidth]{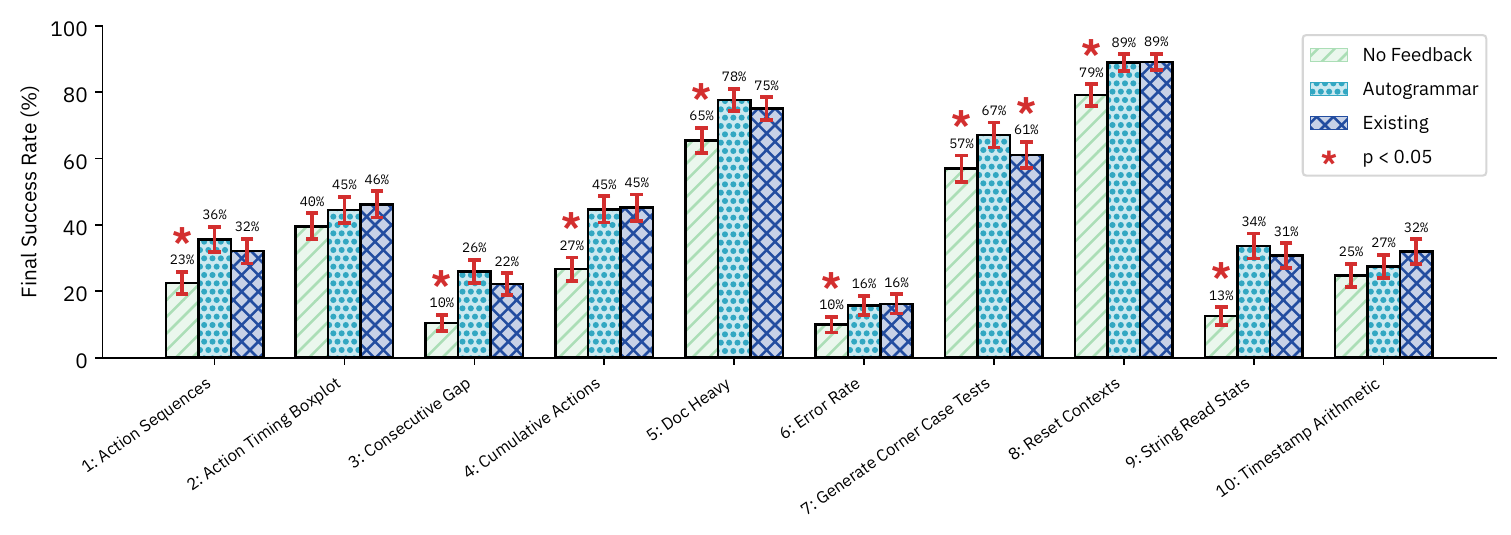}
    \caption{No post-condition versus post-condition using \AG{}-generated grammar 
    versus post-condition using existing grammar. Error bars show 95\% Wilson score 
    confidence intervals. * = $p < 0.05$ (two-proportion z-test).}
    \label{fig:rq4}
\end{figure*}

Figure~\ref{fig:rq4} shows that \AG{} can generate context-free grammars that
improve end-to-end LM performance. Specifically, each bar in
Figure~\ref{fig:rq4} represents the percentage of boto3 CloudWatch Logs Insights
query calls that succeed per task (since there are six LMs and we
call each LM 100 times per task, each bar represents 600 boto3
calls). The bars labeled ``No Feedback'' represent the performance of the first
AI function described above: the baseline performance without any formal
feedback. The bars labeled ``\AG{}'' represent the performance of the AI 
function where \texttt{grammar\_check} uses the context-free grammar generated 
by \AG{}; the bars labeled ``Existing'' represent the performance of the AI 
function where \texttt{grammar\_check} uses the existing CloudWatch 
context-free grammar.

\textit{For 8/10 tasks, \AG{} significantly improves performance over
the baseline. For 1/10 tasks, \AG{} significantly improves the
performance over the professionally maintained grammar,
matching its performance, otherwise.}

\subsection{Threats to Validity}

Our experiments show that \AG{} can learn useful context-free grammars. However,
these results may not generalize due to bias in the log data or tasks that we use.
If our data does not exercise the full breadth
of a target DSL, then \AG{} might only learn a subset of the target language. We mitigate this threat by collecting as
much data as possible and evaluating the
end-to-end performance on a new set of tasks.
If \AG{} were not learning general
context-free grammars, then we would expect to see weaker performance for the
independent experiment used to answer RQ5. To make sure that the tasks we created are representative of real world usage, we
instrumented \AG{} to generate CloudWatch logs, and then we asked real questions
about the logs that we used for analysis in the empirical evaluation of this paper. While the tasks will not represent all agent usage, they do
represent a meaningful case study.
\section{Related Work}
\label{sec:related}

Grammar-constrained decoding is a mature technique. The term was first coined by \citet{geng2023gcd}, who show that
off-the-shelf language models with input-dependent grammars match
task-specific fine-tuned models on structured NLP tasks. Earlier work
used related techniques to enforce syntactic and semantic validity during decoding \cite{scholak2021picard,poesia2022synchromesh}. More recent work, like \citet{willard2023efficient,beurerkellner2024domino}, focus on efficiency. \citet{park2024gad} show that naive
grammar-constrained decoding
distorts the underlying distribution and give an algorithm that recovers
samples proportional to the language model distribution conditioned on the
grammar. All these papers depend on syntactic constraints. In that sense, \citet{wang2023grammarprompting} is the most similar to our work. For each input, they ask LMs to predict a specialized grammar that guides
DSL code generation. \AG{} takes this point of view to an extreme scale by automatically learning a formal artifact that can be independently checked and reused by existing techniques. This is related to skill and experience learning, like \citet{wang2024voyager} and \citet{packer2024memgpt}.

Learning grammars from data has a long history \cite{de2010grammatical}. From a theoretical perspective, \citet{gold1967language}
established that context-free languages are not identifiable from positive data
alone, motivating \AG{}'s use of both positive and negative examples. \citet{angluin1987learning} and \citet{clark2007polynomial} identified fragments of the problem with better complexity. From a practical perspective, GLADE~\cite{bastani2017glade}, Arvada~\cite{kulkarni2022arvada}, TreeVada~\cite{arefin2024treevada}, and Mimid~\cite{gopinath2020mimid} are successful techniques that require oracles or parser instrumentation. \AG{} does not depend on either. Recent work has used LMs to learn grammars \cite{xia2026doc2spec,tang2025hygenar} but focused on documentation or worked on a much smaller data scale. \citet{albinhassan2025learning} tackle a similar problem for context-sensitive constraints, but their evaluation focuses on synthetic languages at a smaller scale, while \AG{} targets real programming languages. The broadest instance of LMs producing formal artifacts is autoformalization, which translates informal input into formal statements, like context-free grammars. \citet{wu2022autoformalization,azerbayev2023proofnet,mora2024speac,polgreen2022uclid5,cosler2023nl2spec,mendoza2026artemis} all fit in this category, among many others. Specification mining---learning formal artifacts from execution data---is also related \cite{ammons2002mining,ernst2007daikon,le2018deepmining}. 

Finally, a growing body of work imposes formal constraints on language model agents. 
For example, \citet{winston2026solver} translate natural-language tool-use policies into constraints and then use these constraints to reject undesirable actions during execution; and \citet{ramani2025bridging} translate natural-language
plans into Kripke structures and then check the plans using a model checker. Both of these approaches are similar in spirit to our work but differ in that we provide a programming framework to define agents as Kripke structures. This opens the door to more powerful tools, like highly expressive constraints over compositions of agents, which we intend to further explore in future work.

\bibliography{main}

\end{document}